\documentclass[12pt]{article}
\pdfoutput=1
\usepackage{draft,tikz-cd}
\usepackage{verbatim}
\usepackage{amsfonts}
\usepackage{mathrsfs}
\usepackage{amsmath}
\usepackage{amssymb}
\usepackage{framed}
\usepackage[medium]{titlesec}
\usepackage{bm}
\usepackage{cite}
\usepackage{soul}
\usepackage[normalem]{ulem}
\usepackage{extarrows}
\usepackage{slashed}
\usepackage{isodateo}
\usepackage{graphicx}
\usepackage{xcolor}
\usepackage[bookmarksnumbered=true,bookmarksopen=true,hyperfootnotes=true]{hyperref}
\hypersetup{colorlinks,%
	linkcolor=[rgb]{0,0.3,0.6}, %
	citecolor=[rgb]{0,0.3,0.6}, %
	urlcolor=[rgb]{0,0.3,0.6}}
\usepackage[hmargin=.7in,vmargin=1.1in]{geometry}
\usepackage{indentfirst}
\usepackage{booktabs}

\usepackage{enumerate}
\usepackage{amsfonts}
\usepackage{mathrsfs}
\usepackage{amsmath}
\usepackage{amssymb}
\usepackage{float}
\usepackage{bm}
\usepackage{slashed}
\usepackage{feynmp-auto}
\usepackage[normalem]{ulem}
\usepackage{extarrows}
\usepackage{slashed}
\usepackage{isodateo}
\usepackage{graphicx}
\usepackage{xcolor}
\usepackage{cancel}
\def\ma{\mathcal}
\def\mb{\mathbf}
\def\te{\text}

\usepackage{feynmp-auto}
\usepackage[normalem]{ulem}
\usepackage{extarrows}
\usepackage{slashed}
\usepackage{isodateo}
\usepackage{graphicx}
\usepackage{xcolor}
\usepackage{cancel}

\definecolor{bblue}{rgb}{0,0.2,0.6}

\renewcommand{\rm}{\mathrm}

\def\mb{\mathbf}

\usepackage{indentfirst}

\graphicspath{{fig/}}

\usepackage{tikz}
\usetikzlibrary{arrows,shapes}
\usetikzlibrary{trees}
\usetikzlibrary{matrix,arrows} 				
\usetikzlibrary{positioning}				
\usetikzlibrary{calc,through}				
\usetikzlibrary{decorations.pathreplacing}  
\usepackage{pgffor}							

\usetikzlibrary{decorations.pathmorphing}	
\usetikzlibrary{decorations.markings}
\tikzset{
	photon/.style={decorate, decoration={snake}, draw=red},
	electron/.style={draw=blue, postaction={decorate},
		decoration={markings,mark=at position .55 with {\arrow[draw=blue]{>}}}},
	gluon/.style={decorate, draw=black,
		decoration={coil,amplitude=4pt, segment length=4pt}} ,
	vector/.style={decorate, decoration={snake}, draw},
	provector/.style={decorate, decoration={snake,amplitude=2.5pt}, draw},
	antivector/.style={decorate, decoration={snake,amplitude=-2.5pt}, draw},
	fermion/.style={draw=black, postaction={decorate},
		decoration={markings,mark=at position .55 with {\arrow[draw=black]{>}}}},
	fermionbar/.style={draw=black, postaction={decorate},
		decoration={markings,mark=at position .55 with {\arrow[draw=black]{<}}}},
	fermionnoarrow/.style={draw=black},
	fermionnoarrowsoft/.style={draw=blue},
	scalar/.style={dashed,draw=black, postaction={decorate},
		decoration={markings,mark=at position .55 with {\arrow[draw=black]{>}}}},
	scalarbar/.style={dashed,draw=black, postaction={decorate},
		decoration={markings,mark=at position .55 with {\arrow[draw=black]{<}}}},
	scalarnoarrow/.style={dashed,draw=black},
	scalarnoarrowsoft/.style={dashed,draw=blue},
	electron/.style={draw=black, postaction={decorate},
		decoration={markings,mark=at position .55 with {\arrow[draw=black]{>}}}},
	bigvector/.style={decorate, decoration={snake,amplitude=4pt}, draw},
}

\tikzstyle{block} = [draw, rectangle, 
minimum height=3em, minimum width=6em]
\usetikzlibrary{arrows,patterns}

\begin{document}

\begin{titlepage}

\begin{center}

\hfill \\
\hfill \\
\vskip 1cm

\title{Notes on weight-shifting operators and unifying relations for cosmological correlators}

\bigskip

\author{Qi Chen,$^{a}$ Yi-Xiao Tao$^{b}$}

\bigskip

\address{${}^a$Department of Physics, Tsinghua University, Beijing 100084, China}

\address{${}^b$Department of Mathematical Sciences, Tsinghua University, Beijing 100084, China}

\email{chenq20@mails.tsinghua.edu.cn, taoyx21@mails.tsinghua.edu.cn}

\end{center}

\vfill

\begin{abstract}
 We seek the inverse formulas for the cosmological unifying relation between gluons and conformally coupled scalars. We demonstrate that the weight-shifting operators derived from the conformal symmetry at the dS late-time boundary can serve as the inverse operators for the 3-point cosmological correlators. However, in the case of the 4-point cosmological correlator, we observe that the inverse of the unifying relation cannot be constructed from the weight-shifting operators. Despite this failure, we are inspired to propose a ``weight-shifting uplifting" method for the 4-point gluon correlator.
\end{abstract}

\vfill

\end{titlepage}

\tableofcontents
\newpage

\section{Introduction}
Cosmic inflation is widely accepted as the period of exponential expansion that occurred at the beginning of our universe, during which the background spacetime can be approximated as de Sitter (dS) space. In the standard inflationary scenario, the early universe is dominated by dark energy, represented by the potential energy of a scalar field known as the inflaton. Quantum fluctuations of the inflaton and other particles during this primordial epoch give rise to non-Gaussianity (NG) and serve as the seeds for the formation of the Large Scale Structure (LSS) we observe today. The NG and LSS contain valuable information that allows us to study the history of the early universe. Extracting such intriguing information is possible through the analysis of cosmological correlation functions \cite{Maldacena:2002vr,Arkani-Hamed:2015bza}. 

Calculating cosmological correlators can be a daunting task. The conventional approach involves employing the in-in formalism to track the time evolution of fields during inflation. This requires integrating over some field mode functions and interaction vertices in dS spacetime, which can be highly intricate. In particular, evaluating the interaction vertices for spinning fields can be arduous, even in flat spacetime, let alone in dS spacetime. However, the in-in formalism provides us with the late-time correlation functions at the future boundary. Motivated by this observation, the cosmological bootstrapping program \cite{Arkani-Hamed:2018kmz, Baumann:2019oyu, Baumann:2020dch, Sleight:2019hfp, DiPietro:2021sjt, Hogervorst:2021uvp, Jazayeri:2021fvk, Baumann:2022jpr, Baumann:2021fxj, Melville:2021lst, Sleight:2021plv} has been proposed. This program aims to compute cosmological correlators from a boundary perspective, leveraging symmetries, locality, unitarity, and other principles. By exploiting these fundamental properties, the cosmological bootstrapping program provides an alternative framework for studying cosmological correlators.

From the inflation perspective, we regard the early universe as a dS$_{4}$ spacetime, which has the same symmetry as CFT$_{3}$. Then we can use many CFT methods to derive these correlators from the conformal properties of the cosmological correlators. In the past decade, people have constructed a series of differential operators which can change the quantum numbers of the operators in cosmological correlators by some CFT methods. These operators are the so-called ``weight-shifting operators" \cite{Baumann:2019oyu,Baumann:2020dch,Bzowski:2022rlz}, and the appearance of these operators allows us to obtain correlators with different quantum numbers from a given cosmological correlator. A brief review of weight-shifting operators is given in section \ref{sec2}.

However, recently the authors find another relation between YM theory and the conformally coupled scalar theory with a gluon minimal coupled. This relation is a generalization of the ``unifying relation" \cite{Cheung:2017ems} for the flat amplitudes. Hence we also use ``unifying relation" to denote this relation for cosmological correlators. The proof, which uses the Berends-Giele recursion, can be found in \cite{Tao:2022nqc}. A generalization to the loop integrand is also found in \cite{Chen:2023bji}.

It seems that there are some connections between these two kinds of relations. In this paper, we want to seek the inverse of unifying relations. In the 3-pt case, we successfully obtain the inverse, while in the 4-pt case things go wrong. To find in what sense we can invert the unifying relations, we construct the 4-pt gluon correlator $\la JJJJ\ra$ from the unifying relations and express it by weight-shifting operators to find some clues. And finally we find that there is a correspondence between the flat amplitudes and the weight-shifting operators in this case. We give a dictionary between the varieties in the flat 4-pt YM amplitudes and the weight-shifting operators that construct $\la JJJJ\ra$.

Our paper will organize as follows. In section \ref{sec2}, we will review some weight-shifting operators and point out the effect they have. In section \ref{sec3}, we will review the unifying relations for cosmological correlators briefly. In section \ref{sec4}, we will construct the 4-pt gluon cosmological correlator $\la JJJJ\ra$ from the unifying relation, and point out that our construction is also consistent with the weight-shifting perspective. Moreover, we point out that this construction can be related to the flat amplitudes.

\section{Weight-shifting operators}\label{sec2}
An interesting way to get the cosmological correlators with correct quantum numbers from other correlators has been introduced in \cite{Arkani-Hamed:2018kmz,Baumann:2019oyu,Baumann:2020dch}, which is known as the ``weight-shifting operators". These operators can be derived from the conformal properties of the cosmological correlators in the embedding space. In this section, we will briefly review these operators and write down some typical examples, following the notation in \cite{Baumann:2020dch}.

A particle in dS spacetime can be labeled by its conformal dimension $\Delta$ and its spin $l$ from the representation of the isometry group of dS spacetime. In cosmology, we only focus on the case dS$_{4}$, which means that we can specialize the weight-shifting operators for dS$_{d+1}$ to the $d=3$ case. The weight-shifting operators are some bi-local operators which can be used to change the conformal dimension $\Delta$ and the spin $l$ of some particles in the correlators, and after acting such operators on a certain correlator we will get another correlator with different quantum numbers. We should point out that we can obtain the correlators with the same quantum numbers by acting on different correlators, and the results may be different. The correct correlators should be a linear combination of these different results. Moreover, sometimes we will get the correlators with wrong singularities, which means that we should do some modifications by hand. Such an example will be given in section \ref{sec4}.


In this section, we only present the explicit expressions for the weight-shifting operators. The derivation and detailed examples illustrating the action of weight-shifting operators on specific correlators will be provided in Appendix \ref{appA}. Let us begin by introducing the simplest differential operators that decrease the conformal dimensions of two fields by one unit in correlators:
\begin{equation}\label{eq:weightlowering12}
\mathcal{W}^{--}_{12}=\frac{1}{2}\vec{K}_{12}\cdot\vec{K}_{12},~~~~~~\vec{K}_{12}=\partial_{\vec{k}_1}-\partial_{\vec{k}_2}.
\end{equation}
In the above expression, the subscript $1,2$ indicates the particle for which we aim to decrease its conformal weight. It is important to emphasize that $\vec{K}_{12}$ represents a differential operator with respect to the component of momentum, and thus, it is a vector. 

Another valuable operator is the one that increases the conformal dimensions of two fields by one unit. The general expression for such weight-raising operators, acting on a spinning field, can be quite intricate. However, for our current purpose, we will not delve further into the details of these operators acting on spinning fields. Instead, we will focus on presenting the expression for the weight-raising operators that specifically act on pure scalar operators:
\ie
\mathcal{W}^{++}_{12}=\frac{1}{2}(k_{1}k_{2})^{2}K_{12}^{2}-(3-2\Delta_{1})(3-2\Delta_{2})\vec{k}_{1}\cdot\vec{k}_{2}+[k_{2}^{2}(3-2\Delta_{1})(2-\Delta_{1}+\vec{k}_{1}\cdot\vec{K}_{12})+(1\leftrightarrow2)].
\fe
For a comprehensive understanding of the general weight-raising operators, one can refer to \cite{Baumann:2019oyu}.


In the preceding discussion, we introduced operators that can increase or decrease the conformal weight. It is worth noting that the 3-point correlation function with one particle having spin can also be determined up to an overall constant through the application of conformal symmetry. Furthermore, the expression for these spinning 3-point correlators exhibits a similar structure to that of scalar fields. Therefore, we can also generate correlators for spinning fields by utilizing differential operators on the scalar 3-point correlator. In other words, we can construct weight-shifting operators that enable us to raise the spins of two fields by one unit within correlators:
\ie\label{eq:S++}
S^{++}_{12}=&(l_{1}+\Delta_{1}-1)(l_{2}+\Delta_{2}-2)\vec{z}_{1}\cdot\vec{z}_{2}-\frac{1}{2}(\vec{z}_{1}\cdot \vec{k}_{1})(\vec{z}_{2}\cdot \vec{k}_{2})K_{12}^{2}\\
&+[(l_{1}+\Delta_{1}-1)(\vec{k}_{2}\cdot\vec{z}_{2})(\vec{z}_{1}\cdot\vec{K}_{12})+(1\leftrightarrow2)].
\fe
Here, $\vec{z}$ represents an auxiliary null vector that coincides with the spin polarization vector $\vec{\epsilon}$ (with $\vec{\epsilon}_{i}\cdot \vec{k}_{i}=0$) in the ``on-shell limit" and this limit can only be taken after all operators have acted. More about this limit can be seen in \cite{Lee:2022fgr}. 

We can also construct weight-shifting operators that simultaneously change the conformal weight and spin of the particles. There are three types of such differential operators, each with a distinct effect. The first type can be represented as
\begin{equation}
    H_{12} = 2(\vec{z}_{1}\cdot\vec{K}_{12})(\vec{z}_{2}\cdot\vec{K}_{12})-(\vec{z}_{1}\cdot\vec{z}_{2})K_{12}^{2}.
\end{equation}
This operator has the capability to raise the spins at points 1 and 2 by one unit while simultaneously lowering their conformal dimensions by one unit. The remaining two types of operators can also simultaneously change the weight and spin of the particles. However, unlike $H_{12}$, they have distinct effects on different particles:
\begin{align}
D_{12}=&(\Delta_{1}+l_{1}-1)(\vec{z}_{1}\cdot\vec{K}_{12})-\frac{1}{2}(\vec{z}_{1}\cdot\vec{k}_{1})K_{12}^{2},\label{eq:Dij}\\
D_{11}=&(\Delta_{2}-3+\vec{k}_{2}\cdot\vec{K}_{12})(\vec{z}_{1}\cdot\vec{K}_{12})-\frac{1}{2}(\vec{z}_{1}\cdot\vec{k}_{2})K_{12}^{2}-(\vec{z}_{2}\cdot\vec{K}_{12})(\vec{z}_{1}\cdot\partial_{\vec{z}_{2}})+(\vec{z}_{1}\cdot\vec{z}_{2})(\partial_{\vec{z}_{2}}\cdot\vec{K}_{12}).
\end{align}
Here, the operator $D_{12}$ raises the spin by one unit at point 1 and lowers the conformal dimension by one unit at point 2. On the other hand, the operator $D_{11}$ raises the spin by one unit at point 1 and lowers the conformal dimension by one unit at the same point.

In the case of conserved spin-1 operators with dimension $\Delta=2$, there is a convenient shortcut. We can treat these conserved spin-1 operators as projections of spin-1 operators with $\Delta=1$. After some algebraic manipulations (as detailed in \cite{Baumann:2020dch}), we find that multiplying the norm of the momentum of the spin-1 currents with $\Delta=1$ gives us the conserved currents. In fact, this operation is also valid for $\Delta=1$ scalars and we will give some examples later. Interestingly, this operation effectively performs the shadow transform. While similar results exist for the spin-2 case, we will not delve into it within the scope of this paper.

Let us write down some results here:
\ie
\la J_{1}\varphi_{2}\varphi_{3}\ra&=D_{11}\la\phi_{1}\varphi_{2}\varphi_{3}\ra,\\
\la J_{1}\varphi_{2}\varphi_{3}\varphi_{4}\ra_{s}&=k_{2}D_{12}\la\phi_1\varphi_2\varphi_{3}\varphi_{4}\ra_{s}.
\fe
where $\phi$ denotes massless scalars, $\varphi$ denotes conformally coupling scalars, and $J$ denotes the conserved currents (contracted with the auxiliary null vectors $\vec{z}$). We need to point out that weight-shifting operators will not change the type of channels, which means that we can act these operators on the correlators with a given channel. The label $s$ in the above example denotes $s$-channel.

Up to this point, our focus has been on differential operators acting on the external legs. However, it is worth noting that we can also construct similar differential operators to change the conformal weight and spin of internal line particles. In the subsequent discussion, we will introduce these differential operators that raise the spin of the internal line particles.

In general, the 4-point correlation function can be decomposed into connected and disconnected parts. The disconnected part refers to the component that can be factorized into the product of two 3-point correlators. This implies that we can construct spin-raising operators for the internal line particles in the disconnected part by utilizing this factorization, as the internal lines in the 4-point correlator become the external legs in the 3-point correlators. As an example, let us consider the disconnected contribution to the 4-point function arising from spin-1 exchange (e.g. $s$-channel):
\begin{equation}
\langle\varphi_{\vec{k}_1}\varphi_{\vec{k}_2}\varphi_{\vec{k}_3}\varphi_{\vec{k}_4}\rangle^{(1)}_{\te{d}}=\frac{\langle\varphi_{\vec{k}_1}\varphi_{\vec{k}_2}\ma{O}^i_{-\vec{s}}\rangle(\Pi_1)_{ij}\langle\ma{O}^j_{\vec{s}}\varphi_{\vec{k}_3}\varphi_{\vec{k}_4}\rangle}{\langle\ma{O}_{\vec{s}}\ma{O}_{-\vec{s}}\rangle},
\end{equation}
where $\ma{O}$ is a scalar operator with dimension $\Delta$, $\vec{s}=\vec{k}_1+\vec{k}_2$ is the momentum of the exchanged particle, and $(\Pi_1)_{ij}$ is a symmetric traceless tensor:
\ie
(\Pi_1)_{ij}=\delta_{ij}+\frac{3-2\Delta}{\Delta-2}\frac{s_{i}s_{j}}{s^2},
\fe
which encodes the polarization structure of the inverse two-point function of the exchanged field: $\langle\ma{O}^i_{\vec{s}}\ma{O}^j_{-\vec{s}}\rangle^{-1}\propto (\Pi_1)_{ij}\langle\ma{O}_{\vec{s}}\ma{O}_{-\vec{s}}\rangle^{-1}$. Then, we can generate the 3-point correlator $\langle\varphi_{\vec{k}_1}\varphi_{\vec{k}2}\ma{O}^i_{-\vec{s}}\rangle$ from the 3-point scalar correlator using the spin-raising operator defined in \eqref{eq:Dij}:

\begin{equation}
    \langle\varphi_{\vec{k}_1}\varphi_{\vec{k}_2}\ma{O}^i_{-\vec{s}}\rangle=\ma{S}^i_{12}\langle\varphi_{\vec{k}_1}\varphi_{\vec{k}_2}\ma{O}_{\vec{s}}\rangle.
\end{equation}
where $\ma{S}_{12}=k_2D_{32}=\sum_{i}\epsilon_{3}^{i}\ma{S}^{i}_{12}$ is the differential operator that raises the spin of the $\ma{O}$ field by one unit. The index $i$ indicates that we have removed the auxiliary null vector $\vec{z}$ from $D_{32}$. Therefore, the disconnected contribution to the 4-point function from spin-1 exchange can be expressed as:
\begin{equation}
\langle\varphi_{\vec{k}_1}\varphi_{\vec{k}_2}\varphi_{\vec{k}_3}\varphi_{\vec{k}_4}\rangle^{(1)}_{\te{d}}=\frac{\ma{S}^i_{12}\langle\varphi_{\vec{k}_1}\varphi_{\vec{k}_2}\ma{O}_{-\vec{s}}\rangle(\Pi_1)_{ij}\ma{S}^j_{34}\langle\ma{O}_{\vec{s}}\varphi_{\vec{k}_3}\varphi_{\vec{k}_4}\rangle}{\langle\ma{O}_{\vec{s}}\ma{O}_{-\vec{s}}\rangle}.
\end{equation}
It is evident that the combination $\ma{S}^i_{12}(\Pi_1)_{ij}\ma{S}^j_{34}$ appearing in the numerator raises the spin of the exchanged particle. If we can manage to extract this differential operator and apply it to the disconnected 4-point correlator, we will obtain the internal line particle spin-raising operator. This process is analogous to the construction of weight-shifting operators for the external leg particles that we discussed earlier. The detailed extraction process is beyond the scope of this paper, but interested readers can refer to \cite{Baumann:2019oyu} for more information. Here, we present the final result obtained from this extraction:
\begin{equation}\label{eq:Suv}
    \langle\varphi_{\vec{k}_1}\varphi_{\vec{k}_2}\varphi_{\vec{k}_3}\varphi_{\vec{k}_4}\rangle^{(1)}_{\te{d}}=\left(\Pi_{1,1}D_{uv}+\Pi_{1,0}\Delta_u\right)\langle\varphi_{\vec{k}_1}\varphi_{\vec{k}_2}\varphi_{\vec{k}_3}\varphi_{\vec{k}_4}\rangle^{(0)}_{\te{d}},
\end{equation}
where $\Pi_{1,1}$ and $\Pi_{1,0}$ are polarization sums for the $s$-channel (we only focus on this case in the following discussion), and $D_{uv}$ and $\Delta_u$ are differential operators:
\begin{align}
    \Pi_{1,1}=\frac{(k_1^2-k_2^2)(k_3^2-k_4^2)}{s^4}+\frac{(\vec{k}_1-\vec{k}_2)(\vec{k}_3-\vec{k}_4)}{s^2},~~~~~~\Pi_{1,0}=\frac{(k_1-k_2)(k_3-k_4)}{s^2},\label{eq:Pi}\\
    D_{uv}=u^2v^2\partial_u\partial_v,~~~~~~\Delta_u=u^2(1-u^2)\partial_u^2-2u^3\partial_u,~~~~~~u\equiv s/(k_1+k_2),~~~~~~v\equiv s/(k_3+k_4).\label{eq:DDelta}
\end{align}
We introduce the dimensionless variables $u$ and $v$ to simplify our writing. The 4-point correlation function $\langle\varphi_{\vec{k}_1}\varphi_{\vec{k}_2}\varphi_{\vec{k}_3}\varphi_{\vec{k}_4}\rangle^{(0)}_{\te{d}}$ can be expressed in terms of these two independent variables $u$ and $v$ due to conformal symmetry ($s$-channel). Thus, such a choice of variables allows us to simplify our calculations in the subsequent sections.

So far, we have focused on the disconnected contribution to the four-point function, considering the exchange of a single scalar operator with a general conformal dimension $\Delta$. However, it is important to note that the generalization to arbitrary spin exchange is straightforward. The disconnected part of the four-point correlator with the arbitrary spin exchange can also be factorized into a product of three-point correlators. Furthermore, it is worth mentioning that the external line spin-raising operator \eqref{eq:Suv} does not depend on the conformal dimension. As a result, we can apply this operator to the full four-point pure scalar correlator, generating a complete four-point scalar correlator with an exchange of spinning fields.

\section{Unifying relations for cosmological correlators}\label{sec3}
The unifying relations\cite{Cheung:2017ems} are first discovered in flat spacetime and some of them are found to be equivalent to the dimensional reduction in Cachazo-He-Yuan (CHY) formalism \cite{Cachazo:2014xea}. Recently, new relations among cosmological correlators have been discovered using Berends-Giele currents \cite{Armstrong:2022mfr,Tao:2022nqc,Chen:2023bji}. These relations can be seen as a generalization of the unifying relations found for flat amplitudes \cite{Cheung:2017ems}, which is why they are also referred to as ``unifying relations" for cosmological correlators. In general, we expect that the entire set of unifying relations from \cite{Cheung:2017ems} can be extended to curved spacetime. However, currently, only the relations involving pure gluons and conformally coupled scalars with minimal gluon coupling have been well established (for the general Yang-Mills-scalar theory, there is also a proof based on BG currents \cite{Chen:2023bji} which can be generalized to the dS case easily). Therefore, in this work, we will focus solely on the case of pure gluon theory and conformally coupled scalars with minimal gluon coupling.

It is important to mention that in the following discussion, we will present the unifying relations without delving into the details of their proofs. We will focus on cosmological correlators involving pure gluons and conformally coupled scalars, and present the following unifying relations:
\ie
\mathcal{T}^{X}A_{\te{YM}}(g_{1},g_{2},\cdots ,g_{n})=A_{\te{S}}(\phi_{X},g_{\{1,2,\cdots,n\}\backslash X}).
\fe
In the given expression, the notation $\phi_{X}$ only points out which particles are scalars, while the order of the particles in $A_{\te{S}}$ is the same as $A_{\te{YM}}$. The differential operator $\mathcal{T}^{X}$ represents the pairing of the letters in the word $X$, resulting in a product of $\mathcal{T}[ij]=\partial_{\epsilon_{i}\cdot\epsilon_{j}}$ (where $\epsilon_{\mu}$ is the polarization vector satisfying the transversal condition). This pairing method is applied to obtain the desired operator, and the final result is obtained by summing over all possible pairing methods. For instance, for the case $X=1234$, the corresponding differential operator is as follows:
\ie
\mathcal{T}^{1234}=\mathcal{T}[12]\mathcal{T}[34]+\mathcal{T}[13]\mathcal{T}[24]+\mathcal{T}[14]\mathcal{T}[23].
\fe
Furthermore, we can explicitly demonstrate the action of $\mathcal{T}[ij]$. Let us consider the 4-point gluon correlator as an example, which allows us to establish a relation between the gluon correlator and correlators involving both gluons and conformally coupled scalars through the operation of $\mathcal{T}[ij]$:
\ie
\mathcal{T}[12]A_{\te{YM}}(g_{1},g_{2},g_{3},g_{4})&=A_{\te{S}}(\phi_{1},\phi_{2},g_{3},g_{4})\\
\mathcal{T}[13]A_{\te{YM}}(g_{1},g_{2},g_{3},g_{4})&=A_{\te{S}}(\phi_{1},g_{2},\phi_{3},g_{4}).
\fe
We will not delve into the general formalism for cosmological unifying relations in detail here, but instead, we will provide explicit examples in the following discussion. For a more comprehensive introduction, readers can refer to \cite{Tao:2022nqc}. However, before we proceed to specific examples, it is important to make some remarks regarding cosmological unifying relations. Unlike the weight-shifting operators discussed in Section \ref{sec2}, which heavily rely on conformal symmetry, the unifying relation operators $\mathcal{T}[ij]$ are in both flat and dS spacetime.

It is worth mentioning that the aforementioned relation remains valid within a specific channel since the operator $\mathcal{T}^{X}$ does not alter the channel. To demonstrate this explicitly, let's consider the explicit expressions for the 4-point correlators involving pure gluons and correlators with both gluons and conformally coupled scalars. These correlators can be computed using Berends-Giele currents \cite{Armstrong:2022mfr} or directly from Feynman diagrams \cite{Kharel:2013mka,Albayrak:2018tam,Albayrak:2019asr}. Here, we will provide the results for these correlators\footnote{Here we use the correlators $\la JJJJ\ra$ in \cite{Baumann:2020dch}, while in \cite{Tao:2022nqc} there is an overall minus sign from some conventions.}:  
\begin{equation}\label{eq:JJJJ}
\begin{split}
    \langle JJJJ\rangle_s=&-\frac{1}{k_{s}^{2}}\frac{(k_1-k_2)(k_3-k_4)}{k_1+k_2+k_3+k_4}(\vec{\epsilon}_1\cdot\vec{\epsilon}_2)(\vec{\epsilon}_3\cdot\vec{\epsilon}_4)+\frac{(k_{1}^{2}-k_{2}^{2})(k_{3}^{2}-k_{4}^{2})(\vec{\epsilon}_{1}\cdot\vec{\epsilon}_{2})(\vec{\epsilon}_{3}\cdot\vec{\epsilon}_{4})}{k_{s}^{2}(k_{1}+k_{2}+k_{s})(k_{3}+k_{4}+k_{s})(k_{1}+k_{2}+k_{3}+k_{4})}\\
    &+\frac{\eta^{ij}[2\epsilon_{1i}(\vec{k}_1\cdot\vec{\epsilon}_2)-k_{1i}(\vec{\epsilon}_1\cdot\vec{\epsilon}_2)-(1\leftrightarrow2)][2\epsilon_{3j}(\vec{k}_3\cdot\vec{\epsilon}_4)-k_{3j}(\vec{\epsilon}_3\cdot\vec{\epsilon}_4)-(3\leftrightarrow4)]}{(k_1+k_2+k_3+k_4)(k_1+k_2+k_s)(k_3+k_4+k_s)},
\end{split}
\end{equation}

\begin{equation}\label{eq:JJOO}
    \begin{split}
    \langle JJ\varphi\varphi\rangle_{s}=&-\frac{(\vec{\epsilon}_{1}\cdot\vec{\epsilon}_{2})(k_{1}-k_{2})(k_{3}-k_{4})}{k_{s}^{2}(k_{1}+k_{2}+k_{3}+k_{4})}+
    \frac{(k_{1}^{2}-k_{2}^{2})(k_{3}^{2}-k_{4}^{2})(\vec{\epsilon}_{1}\cdot\vec{\epsilon}_{2})}{k_{s}^{2}(k_{1}+k_{2}+k_{s})(k_{3}+k_{4}+k_{s})(k_{1}+k_{2}+k_{3}+k_{4})}\\
    &+\frac{(\vec{\epsilon}_{1}\cdot\vec{\epsilon}_{2})(\vec{k}_1-\vec{k}_2)\cdot(\vec{k}_3-\vec{k}_4)+4(\vec{k}_2\cdot\vec{\epsilon}_1)(\vec{k}_3\cdot\vec{\epsilon}_2)-4(\vec{k}_1\cdot\vec{\epsilon}_2)(\vec{k}_3\cdot\vec{\epsilon}_1)}{(k_{1}+k_{2}+k_{s})(k_{3}+k_{4}+k_{s})(k_{1}+k_{2}+k_{3}+k_{4})},
\end{split}
\end{equation}
and
\begin{equation}\label{eq:OOOO}
    \begin{split}
        \langle \varphi\varphi\varphi\varphi\rangle_s=-\frac{1}{k_{s}^{2}}\frac{(k_1-k_2)(k_3-k_4)}{k_1+k_2+k_3+k_4}&+\frac{(k_1^2-k_2^2)(k_3^2-k_4^2)}{k_{s}^{2}(k_1+k_2+k_3+k_4)(k_1+k_2+k_s)(k_3+k_4+k_s)}\\
		&+\frac{(\vec{k}_1-\vec{k}_2)\cdot(\vec{k}_3-\vec{k}_4)}{(k_1+k_2+k_3+k_4)(k_1+k_2+k_s)(k_3+k_4+k_s)}.
    \end{split}
\end{equation}
It is important to note that in the above expressions for $\langle JJJJ\rangle$, $\langle JJ\varphi\varphi\rangle$, and $\langle \varphi\varphi\varphi\varphi\rangle$, we have only presented the $s$-channel contribution and neglected the contact contribution. Here, the symbol $J$ represents the conserved current, while $\varphi$ denotes the conformally coupled scalar with a conformal weight of $\Delta=2$. Additionally, $\eta^{ij}$ represents the late time boundary metric, which coincides with the 3D flat metric with the Euclidean signature.

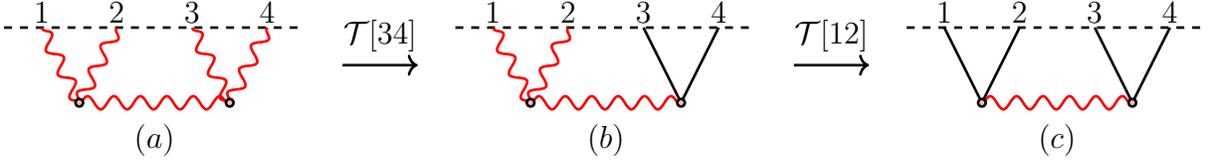
\begin{figure}[t]
\centering
\begin{tikzpicture}[line width=1pt,scale=1]
\draw[scalarnoarrow](0,0)--(4,0);
\draw[photon](1,-1)--(0.5,0);
\draw[photon](3,-1)--(2.5,0);
\draw[photon](1,-1)--(1.5,0);
\draw[photon](3,-1)--(3.5,0);
\draw[photon](1,-1)--(3,-1);
\draw[fermionnoarrow,fill=pink] (1,-1) circle (.05cm);
\draw[fermionnoarrow,fill=pink] (3,-1) circle (.05cm);
\node at (2,-1.5) {$(a)$};
\node at (0.5,0.2) {1};
\node at (1.5,0.2) {2};
\node at (2.5,0.2) {3};
\node at (3.5,0.2) {4};
\node at (5,-0.1) {$\ma{T}[34]$};
\draw[->,line width =1pt](4.5,-0.5)--(5.5,-0.5);
\begin{scope}[shift={(6,0)}]
\draw[scalarnoarrow](0,0)--(4,0);
\draw[photon](1,-1)--(0.5,0);
\draw[fermionnoarrow](3,-1)--(2.5,0);
\draw[photon](1,-1)--(1.5,0);
\draw[fermionnoarrow](3,-1)--(3.5,0);
\draw[photon](1,-1)--(3,-1);
\draw[fermionnoarrow,fill=pink] (1,-1) circle (.05cm);
\draw[fermionnoarrow,fill=pink] (3,-1) circle (.05cm);
\node at (2,-1.5) {$(b)$};
\node at (0.5,0.2) {1};
\node at (1.5,0.2) {2};
\node at (2.5,0.2) {3};
\node at (3.5,0.2) {4};
\node at (5,-0.1) {$\ma{T}[12]$};
\draw[->,line width =1pt](4.5,-0.5)--(5.5,-0.5);
\end{scope}
\begin{scope}[shift={(12,0)}]
\draw[scalarnoarrow](0,0)--(4,0);
\draw[fermionnoarrow](1,-1)--(0.5,0);
\draw[fermionnoarrow](3,-1)--(2.5,0);
\draw[fermionnoarrow](1,-1)--(1.5,0);
\draw[fermionnoarrow](3,-1)--(3.5,0);
\draw[photon](1,-1)--(3,-1);
\draw[fermionnoarrow,fill=pink] (1,-1) circle (.05cm);
\draw[fermionnoarrow,fill=pink] (3,-1) circle (.05cm);
\node at (2,-1.5) {$(c)$};
\node at (0.5,0.2) {1};
\node at (1.5,0.2) {2};
\node at (2.5,0.2) {3};
\node at (3.5,0.2) {4};
\end{scope}
\end{tikzpicture}
\caption{The $s$-channel Witten diagrams for $\langle JJJJ\rangle$, $\langle JJ\varphi\varphi\rangle$, and $\langle \varphi\varphi\varphi\varphi\rangle$. The dashed line represents the late-time boundary of the dS space, the wavy line denotes the gluon propagator, and the solid line denotes the propagator for the conformally coupled scalar. This demonstrates that the correlators $\langle JJ\varphi\varphi\rangle$ and $\langle\varphi\varphi\varphi\varphi\rangle$ can be derived from the 4-point gluon correlators using the unifying relation operators $\mathcal{T}[34]$ and $\mathcal{T}[34]\mathcal{T}[12]$.}\label{fig:unifyingrelation}
\end{figure}

By comparing equations \eqref{eq:JJJJ}, \eqref{eq:JJOO}, and \eqref{eq:OOOO}, we observe that these three correlators can be related through the differential operators $\partial_{\vec{\epsilon}_1\cdot\vec{\epsilon}_2}$ and $\partial_{\vec{\epsilon}_3\cdot\vec{\epsilon}_4}$, which precisely correspond to the unifying relation operators $\ma{T}[12]$ and $\ma{T}[34]$. To provide a more explicit expression (Fig. \ref{fig:unifyingrelation}), we can write these relations as follows:
\ie\label{eq:4-ptunifying}
\ma{T}[12]\ma{T}[34]\langle JJJJ\rangle_s=\ma{T}[12]\langle JJ\varphi\varphi\rangle_{s}=\langle \varphi\varphi\varphi\varphi\rangle_s.
\fe
Here for $\la JJJJ\ra_{s}$, the action of $\ma{T}^{1234}$ is equivalent to the action of $\ma{T}[12]\ma{T}[34]$.
The unifying relation for the 4-point gluon correlator appears to be relatively straightforward, as we can explicitly write down the expression for the correlator with lower-point functions.  However, finding the inverse operators for these unifying relations is an ongoing endeavor and presents a significant challenge. The difficulties arise from the fact that certain terms in $\langle JJJJ\rangle_s$ and $\langle JJ\varphi\varphi\rangle_s$ do not contain $(\vec{\epsilon}_1\cdot\vec{\epsilon}_2)$ or $(\vec{\epsilon}_3\cdot\vec{\epsilon}_4)$. These terms, which do not involve the polarization product $(\vec{\epsilon}_1\cdot\vec{\epsilon}_2)$ or $(\vec{\epsilon}_3\cdot\vec{\epsilon}_4)$, will disappear after applying the unifying relation operators $\ma{T}[12]$ and $\ma{T}[34]$. Therefore, the most difficult issue in seeking the inverse of the unifying relation operators is how to restore these terms. It is possible that certain symmetries may assist in achieving the restoration. Indeed, weight-shifting arising from conformal symmetry may provide insights into restoring these terms and constructing the inverse of the unifying relation operators. Therefore, in the subsequent sections, we aim to find the inverse of the unifying relation operators using the weight-shifting operators.



\section{Seeking the inverse of unifying relations}\label{sec4}

In the previous section, we demonstrated the potential of weight-shifting operators derived from conformal symmetry in constructing the inverse of the unifying relation. In this section, we will provide concrete examples to support this claim. While constructing the inverse, we will also highlight the numerous unexplored weight-shifting methods. We will begin with a discussion of the 3-pt case as a warm-up before delving into the more challenging 4-pt correlators. Unfortunately, we will encounter difficulties in directly constructing the inverse of the unifying relation operators. However, in dS spacetime, there exists the uplifting method \cite{Baumann:2021fxj,Bonifacio:2022vwa,Li:2022tby,Mei:2023jkb} that allows us to obtain dS correlators from flat amplitudes by some replacement. In the final part of this section, we will explore the analogous uplifting method for weight-shifting operators in dS space. It will be interesting to compare these two different uplifting methods.

\subsection{3-pt inverse: warm-up}
Due to the favorable property that we can determine 3-point correlators up to a coupling constant through conformal symmetry, our primary focus lies on exploring the inverse of unifying relation operators for the 3-point correlator.
We anticipate that we can recover all the information encoded by conformal symmetry. To facilitate our discussion, let us begin with an explicit presentation of the 3-point gluon correlators. Similar to the 4-point gluon correlator discussed in Section \ref{sec3}, we will provide the final results here, while referring interested readers to previous papers\cite{Armstrong:2022mfr,Tao:2022nqc} for detailed calculations:
\begin{equation}\label{eq:JJJ}
    \langle JJJ\rangle=\frac{2(\vec{\epsilon}_1\cdot\vec{\epsilon}_2)(\vec{\epsilon}_3\cdot\vec{k}_1)}{k_1+k_2+k_3}+\frac{2(\vec{\epsilon}_2\cdot\vec{\epsilon}_3)(\vec{\epsilon}_1\cdot\vec{k}_2)}{k_1+k_2+k_3}+\frac{2(\vec{\epsilon}_1\cdot\vec{\epsilon}_3)(\vec{\epsilon}_2\cdot\vec{k}_3)}{k_1+k_2+k_3}.
\end{equation}
Next, we can utilize the unifying relation operators to derive the 3-pt mixed correlator between conformally coupled scalars and the conserved current from the 3-pt pure gluon correlators:
\begin{equation}\label{eq:Jvarphivarphi}
    \la J\varphi\varphi\ra=\ma{T}[23]\langle JJJ\rangle=\frac{2(\vec{\epsilon}_{1}\cdot \vec{k}_{2})}{k_{1}+k_{2}+k_{3}}.
\end{equation}
Now, our goal is to find suitable weight-shifting operators that can recover the 3-pt gluon correlators $\langle JJJ\rangle$ in \eqref{eq:JJJ} from $\langle J\varphi\varphi\rangle$ in \eqref{eq:Jvarphivarphi}. There are various methods to generate a correlation function with the correct kinematics, and one possible approach is to utilize the spin-raising operator \eqref{eq:S++}:
\begin{equation}
    \langle JJJ\rangle=S^{++}_{23}\langle J\varphi\varphi\rangle+\te{cycclic permutations}.
\end{equation}
Indeed, utilizing the spin-raising operators allows us to restore the full result for the 3-pt gluon correlator. However, it is important to note that these naive spin-raising operators cannot serve as the inverse of our unifying relation operators. The reason for this is that we need to incorporate all the possible information contained in the 3-pt mixed correlators involving both conformally coupled scalars and gluons, such as $\langle\varphi J\varphi\rangle$ and $\langle\varphi\varphi J\rangle$. The concept of inverse implies that we expect to recover the 3-pt gluon correlator only from $\langle J\varphi\varphi\rangle$, which is not achievable with these spin-raising operators alone.

\begin{figure}[t]
\centering
\begin{tikzpicture}[line width=1pt,scale=1]
\draw[scalarnoarrow](-1,0)--(5,0);
\draw[photon](2,-2)--(2,0);
\draw[photon](2,-2)--(0,0);
\draw[photon](2,-2)--(4,0);
\draw[fermionnoarrow,fill=pink] (2,-2) circle (.05cm);
\node at (2,-2.5) {$(a)$};
\node at (0,0.2) {1};
\node at (2,0.2) {2};
\node at (4,0.2) {3};
\node at (6,-0.4) {$\ma{T}[23]$};
\draw[->,line width =1pt](5,-0.8)--(7,-0.8);
\draw[->,line width =1pt](7,-1.4)--(5,-1.4);
\node at (6,-1.8) {$\ma{T}[23]^{-1}$};
\begin{scope}[shift={(8,0)}]
\draw[scalarnoarrow](-1,0)--(5,0);
\draw[fermionnoarrow](2,-2)--(2,0);
\draw[photon](2,-2)--(0,0);
\draw[fermionnoarrow](2,-2)--(4,0);
\draw[fermionnoarrow,fill=pink] (2,-2) circle (.05cm);
\node at (2,-2.5) {$(b)$};
\node at (0,0.2) {1};
\node at (2,0.2) {2};
\node at (4,0.2) {3};
\end{scope}
\end{tikzpicture}
\caption{The Feynman diagram shown here depicts the 3-pt pure gluon correlator ($a$) and the 3-pt mixed correlator between conformally coupled scalars and gluon ($b$). The mixed correlator can be obtained from the pure gluon correlator through the action of the unifying relation operator $\ma{T}[23]$. Conversely, we can also obtain the pure gluon correlator from the mixed 3-pt correlator by employing $\ma{T}[23]^{-1}$, which is constructed from a set of weight-shifting operators.}\label{fig:3-ptinverse}
\end{figure}
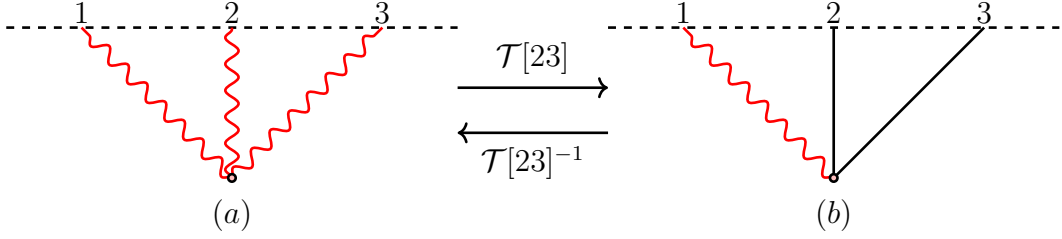

However, it is important to reiterate that the 3-pt correlators are determined only up to a coupling constant. Consequently, there is a possibility to recover the 3-pt gluon correlator only from $\langle J\varphi\varphi\rangle$, which allows us to find out the inverse of the unifying relation operators $\ma{T}[23]$:
\begin{equation}\label{eq:3-ptinverse}
    \langle JJJ\rangle=\ma{T}[23]^{-1}\langle J\varphi\varphi\rangle,~~~~~~\ma{T}[23]^{-1}=(H_{23}+D_{22}D_{33}-2D_{23}D_{32})\ma{W}_{23}^{++}.
\end{equation}
It is evident from the equation \eqref{eq:3-ptinverse} presented that the operator $\ma{T}[23]^{-1}$ can be utilized to restore the 3-pt gluon correlator $\langle JJJ\rangle$ from the mixed correlator $\langle J\varphi\varphi\rangle$. This implies that we have successfully identified the inverse of the unifying relation operator $\ma{T}[23]$. Consequently, we can initiate the process with a 3-pt pure gluon correlator and generate a mixed correlator through the application of a differential operator, and vice versa (see Fig. \ref{fig:3-ptinverse}).

While it is true that the 3-pt correlators are determined up to coupling constants, the situation becomes more intricate when multiple particles possess spin \cite{Simmons-Duffin:2016gjk}. In such cases, there can exist more than one linearly independent structure that complies with conformal invariance. To illustrate, when generating the 3-pt gluon correlator \eqref{eq:JJJ} from the mixed correlator \eqref{eq:Jvarphivarphi} using a weight-shifting operator, one possible approach is as follows:
\begin{equation}
    \langle JJJ\rangle=k_{2}k_{3}H_{23}\langle J\varphi\varphi\rangle.
\end{equation}
This particular weight-shifting operator successfully reproduces the appropriate quantum numbers, including both spin and conformal dimension, for $\langle JJJ\rangle$. However, it introduces additional pole structures that are not expected to be present. In fact, the 3-pt pure gluon correlator obtained using these weight-shifting operators corresponds to the $\mathrm{Tr}(F^3)$ term, where $F$ represents the YM field strength tensor.

\subsection{4-pt inverse: weight-shifting uplifting?}
However, the situation becomes more intricate for higher-point correlators. While conformal invariance can still yield certain structures in 4-point correlators, directly inverting the unifying relation proves to be challenging. For higher-point correlators, conformal invariance does not offer a straightforward solution, suggesting that a readily available weight-shifting prescription may not exist. In the subsequent discussion of this subsection, our focus remains on our attempt at constructing the inverse of unifying relation operators. We want to emphasize once again that we only consider the $s$-channel in this subsection.

Before delving into the formal discussion of the inverse of unifying relation operators in 4-point correlators, let us establish some conventions for this subsection. We denote $F_{\Delta=2}$ as the 4-point scalar correlator with a conformally coupled exchange in the internal line. Additionally, we use $\langle\varphi\varphi\varphi\varphi\rangle$ to represent the 4-point scalar correlator with a spinning particle exchange in the internal line.
\subsubsection{Known results and discussions}
Let us begin by examining the simpler cases involving 4-point mixed correlators between two conserved currents and two conformally coupled scalars. In this discussion, we will not consider scenarios where the internal line corresponds to spinning fields; instead, our focus will be on the cases with scalar exchanges. For instance, consider the correlator $\langle J\varphi J\varphi\rangle$, where the internal line represents a conformally coupled scalar. Drawing inspiration from the 3-point cases, we can employ weight-shifting operators for the external legs to generate $\langle J\varphi J\varphi\rangle$ from $F_{\Delta=2}$ since there is no need to raise the spin of the internal line\footnote{In \cite{Baumann:2020dch}, they omit the overall constant in the eq. (5.24).}:
\ie
\la J\varphi J\varphi\ra=4k_{2}D_{12}k_{4}D_{34}F_{\Delta=2}=\frac{4(\vec{k}_2\cdot\vec{\epsilon}_1)(\vec{k}_4\cdot\vec{\epsilon}_3)}{(k_1+k_2+k_s)(k_3+k_4+k_s)(k_1+k_2+k_3+k_4)}.
\fe
It is important to note that all calculations in our work are performed within the boundary Lorenz gauge ($\eta^{\mu\nu}\partial_{\mu}A_{\nu}=0$). The boundary correlation functions are inherently gauge-invariant, and the weight-shifting operators operate on these gauge-invariant correlators\footnote{It is interesting to point out that the gauge invariance holds for the transverse part of the correlators but the full correlators actually obey non-trivial Ward identities due to contact terms \cite{Armstrong:2020woi}. In this paper, we do not consider such contact terms.}. Consequently, the expressions for the differential operators should remain consistent, even when considered in a different gauge. Certainly, the $s$-channel diagrams results may be influenced by the choice of gauge. However, since the weight-shifting operators will not change the type of channels, they remain gauge-invariant even though they act on a certain channel rather than the total correlator.

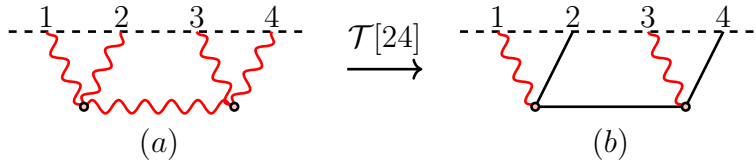
\begin{figure}[t]
\centering
\begin{tikzpicture}[line width=1pt,scale=1]
\draw[scalarnoarrow](0,0)--(4,0);
\draw[photon](1,-1)--(0.5,0);
\draw[photon](3,-1)--(2.5,0);
\draw[photon](1,-1)--(1.5,0);
\draw[photon](3,-1)--(3.5,0);
\draw[photon](1,-1)--(3,-1);
\draw[fermionnoarrow,fill=pink] (1,-1) circle (.05cm);
\draw[fermionnoarrow,fill=pink] (3,-1) circle (.05cm);
\node at (2,-1.5) {$(a)$};
\node at (0.5,0.2) {1};
\node at (1.5,0.2) {2};
\node at (2.5,0.2) {3};
\node at (3.5,0.2) {4};
\node at (5,-0.1) {$\ma{T}[24]$};
\draw[->,line width =1pt](4.5,-0.5)--(5.5,-0.5);
\begin{scope}[shift={(6,0)}]
\draw[scalarnoarrow](0,0)--(4,0);
\draw[photon](1,-1)--(0.5,0);
\draw[photon](3,-1)--(2.5,0);
\draw[fermionnoarrow](1,-1)--(1.5,0);
\draw[fermionnoarrow](3,-1)--(3.5,0);
\draw[fermionnoarrow](1,-1)--(3,-1);
\draw[fermionnoarrow,fill=pink] (1,-1) circle (.05cm);
\draw[fermionnoarrow,fill=pink] (3,-1) circle (.05cm);
\node at (2,-1.5) {$(b)$};
\node at (0.5,0.2) {1};
\node at (1.5,0.2) {2};
\node at (2.5,0.2) {3};
\node at (3.5,0.2) {4};
\end{scope}
\end{tikzpicture}
\caption{The $s$-channel Witten diagrams for $\langle JJJJ\rangle$, $\langle J\varphi J\varphi\rangle$. The dashed line represents the late-time boundary of the dS space, the wavy line denotes the gluon propagator, and the solid line denotes the propagator for the conformally coupled scalar. This shows that the application of $\ma{T}[24]$ results in a simultaneous decrease in the spin of the internal line.}\label{fig:unifyingrelation24}
\end{figure}

There does not exist a known weight-shifting operator to obtain the 4-point gluon correlator $\langle JJJJ\rangle$ from $\langle J\varphi J\varphi\rangle$. This outcome may not come as a surprise because $\ma{T}[24]$ is known to decrease both the spin of the internal line and the external legs (Fig. \ref{fig:unifyingrelation24}), and the operators from the 3-point case are required to act on two points connected to the same vertex. Conversely, in order to find the inverse of $\ma{T}[24]$, we would need to identify an operator that can increase the spin of both the internal line and the external legs. However, to find this operator we need to regard $\la J\varphi J\varphi\ra$ as a product of two 3-pt correlators and then act operators on these two 3-pt correlators, and then we will find a singularity from the polarization tensor $(\Pi_{1})_{ij}$. In the next subsection, we will discuss more about this operation.

Next, we turn our attention to the more intricate scenario where the 4-point mixed correlators involve spinning field exchanges in the internal line. Consider, for instance, the correlator $\langle JJ\varphi\varphi\rangle$, which features a spinning field exchange in the internal line. Recall that in the case of the 3-point correlators, we successfully constructed the inverse of the unifying relation operators \eqref{eq:3-ptinverse}, enabling us to generate $\langle JJJ\rangle$ from $\langle J\varphi\varphi\rangle$. In the current situation, we also need to raise the spin of two external legs, suggesting the feasibility of considering the inverse of the unifying relation operator in \eqref{eq:3-ptinverse}:
\ie\label{eq:JJvarphivarphiincorrect}
\langle JJ\varphi\varphi\rangle\sim(H_{12}+D_{11}D_{22}-2D_{12}D_{21})\ma{W}_{12}^{++}\la\varphi\varphi\varphi\varphi\ra.
\fe
Although this combination of weight-shifting operators preserves the correct quantum numbers, it does not yield the expected expression for $\langle JJ\varphi\varphi\rangle$ in the case of 4-point correlators. Unlike what we observed in the 3-point correlators, the combination of weight-shifting operators in \eqref{eq:JJvarphivarphiincorrect} leads to different pole structures compared to \eqref{eq:JJOO}. Specifically, the result in \eqref{eq:JJvarphivarphiincorrect} exhibits the same pole structure as the contact term rather than the desired poles. To explicitly reveal its pole structure, we can express $\langle\varphi\varphi\varphi\varphi\rangle$ in terms of $F_{\Delta=2}$ using the spin-raising operator for the internal line in \eqref{eq:Suv}:
\ie\label{eq:JJvarphivarphiincorrect1}
(H_{12}+D_{11}D_{22}-2D_{12}D_{21})\ma{W}_{12}^{++}\la\varphi\varphi\varphi\varphi\ra=[\ma{S}_{12}^{++}(\Pi_{1,1}D_{uv}+\Pi_{1,0}\Delta_{u})+2(\vec{\epsilon}_{1}\circ\vec{\epsilon}_{2})D_{uv}]\Delta_{u}(\Delta_{u}-12)F_{\Delta=2}.
\fe
Here, $S^{++}_{12}$ represents the spin-raising operator for external legs, $\Pi_{1,1}$ and $\Pi_{1,0}$ denote the spin polarization sums introduced in \eqref{eq:Pi}. Furthermore, $D_{uv}$ and $\Delta_u$ are differential operators with respect to dimensionless variables $u$ and $v$, as defined in \eqref{eq:DDelta}. To simplify notation, we introduce the circle product between two polarization vectors $\vec{\epsilon}$, which is defined as:
\begin{equation}
    \vec{\epsilon}_1\circ\vec{\epsilon}_2=\frac{2}{k_s^2}\left[(\vec{\epsilon}_1\cdot\vec{k}_3)(\vec{\epsilon}_2\cdot\vec{k}_4)-(\vec{\epsilon}_1\cdot\vec{k}_4)(\vec{\epsilon}_2\cdot\vec{k}_3)\right].
\end{equation}
It is worth noting that the singularity of \eqref{eq:JJvarphivarphiincorrect1} is not correct for the $s$-channel part of a 4-pt correlator. However, the presence of the additional factor $\Delta_u(\Delta_u-12)$ can give the right singularity. In order to eliminate these extra poles and obtain the correct result for $\langle JJ\varphi\varphi\rangle$, we need to manually replace $\Delta_u(\Delta_u-12)F_{\Delta=2}$ with $F_{\Delta=2}$:
\begin{equation}\label{eq:JJvarphivarphi}
    \langle JJ\varphi\varphi\rangle=[S_{12}^{++}(\Pi_{1,1}D_{uv}+\Pi_{1,0}\Delta_{u})+2(\vec{\epsilon}_{1}\circ\vec{\epsilon}_{2})D_{uv}]F_{\Delta=2}.
\end{equation}

We should bear in mind that the 4-pt mixed correlator $\langle JJ\varphi\varphi\rangle$ and the 4-pt pure scalar correlator with a spinning field exchange in the internal line are related through the unifying relation operators $\ma{T}[12]$, as demonstrated in \eqref{eq:4-ptunifying}. However, it seems important to construct a combination of weight-shifting operators that can serve as the inverse of the unifying relation operator $\ma{T}[12]$ based on the weight-shifting operator prescription without the need for any manual replacement procedure. Such a failure, together with the failure of the $\la J\varphi J\varphi\ra$ case, can be understood since the conformal symmetry cannot determine the structure of 4-pt correlators, which means that there will still be some information that cannot be restored. In the following discussion, we will show how to construct $\la JJJJ\ra$ in the weight-shifting meaning, and then show the hints of a weight-shifting uplifting method for cosmological correlators.

\subsubsection{$\la JJJJ\ra$ and hints of a weight-shifting uplifting method}
The unifying relation tells us that for $\la JJJJ\ra$, where all terms include a $\vec{\epsilon}_i\cdot\vec{\epsilon}_j$, we can multiply the corresponding $\vec{\epsilon}_i\cdot\vec{\epsilon}_j$ to $\la JOJO\ra$, $\la JJOO\ra$, and so on. From this prescription, we can find the following result:
\ie
\la JJJJ\ra=&(\vec{\epsilon}_1\cdot\vec{\epsilon}_2)\la\varphi\varphi JJ\ra+(\vec{\epsilon}_3\cdot\vec{\epsilon}_4)\la JJ\varphi\varphi\ra-(\vec{\epsilon}_1\cdot\vec{\epsilon}_2)(\vec{\epsilon}_3\cdot\vec{\epsilon}_4)\la\varphi\varphi\varphi\varphi\ra\\
&+(\vec{\epsilon}_{1}\cdot\vec{\epsilon}_{3})\la \varphi J\varphi J\ra+(\vec{\epsilon}_{1}\cdot\vec{\epsilon}_{4})\la \varphi JJ\varphi\ra+(\vec{\epsilon}_{2}\cdot\vec{\epsilon}_{3})\la J\varphi\varphi J\ra+(\vec{\epsilon}_{2}\cdot\vec{\epsilon}_{4})\la J\varphi J\varphi\ra.
\fe
For terms proportional to $(\vec{\epsilon}_1\cdot\vec{\epsilon}_2)(\vec{\epsilon}_3\cdot\vec{\epsilon}_4)$ in $\la JJJJ\ra$, they will  be contributed by both $(\vec{\epsilon}_1\cdot\vec{\epsilon}_2)\la\varphi\varphi JJ\ra$ and $(\vec{\epsilon}_3\cdot\vec{\epsilon}_4)\la JJ\varphi\varphi\ra$. Therefore, we must add the term $-(\vec{\epsilon}_1\cdot\vec{\epsilon}_2)(\vec{\epsilon}_3\cdot\vec{\epsilon}_4)\la\varphi\varphi\varphi\varphi\ra$ to cancel the repeated part. This equation satisfies the unifying relation manifestly. We can also use the seed function $F_{\Delta=2}$ to get a more compact formalism:
\begin{equation}\label{jjjj}
    \begin{split}
\langle JJJJ\rangle=&[S^{++}_{12}S^{++}_{34}(\Pi_{1,1}D_{uv}+\Pi_{1,0}\Delta_{w})+2S^{++}_{34}(\vec{\epsilon}_{1}\circ\vec{\epsilon}_{2})D_{uv}+2S^{++}_{12}(\vec{\epsilon}_{3}\circ\vec{\epsilon}_{4})D_{uv}\\
&+4S^{++}_{24}k_{4}D_{34}k_{2}D_{12}+4S^{++}_{13}k_{3}D_{43}k_{1}D_{21}-4S^{++}_{23}k_{2}k_{3}D_{12}D_{43}-4S^{++}_{14}k_{1}k_{4}D_{21}D_{34}]F_{\Delta=2}.
\end{split}
\end{equation}
This formalism coincides with the statement in \cite{Lee:2022fgr}, that the differential operators in unifying relations can be written as functional derivatives with respect to the weight-shifting operators. Note that the order of the weight-shifting operators is the same as the order defined in \cite{Lee:2022fgr}. However, this equation seems not consistent with the weight-shifting perspective. The first line of the equation has raised the spin of the internal line, while the second line of the equation has not. From a weight-shifting perspective, we may find that the second line of the equation corresponds to a scalar internal line, which is not consistent with $\la JJJJ\ra$. This inconsistency, however, does not exist. In the following discussion, we will show that the second line of the equation can also be obtained from a weight-shifting process with the spin of the internal line being raised.


$\la J\varphi\varphi J\ra$ can be written as a product of two 3-pt correlators, which is the characterization of the disconnected part of 4-pt correlators:
\ie
\la J\varphi\varphi J\ra=\frac{-4(\vec{\epsilon}_{1}\cdot \vec{k}_{2})(\vec{\epsilon}_{4}\cdot \vec{k}_{3})}{E(k_{1}+k_{2}+k_{s})(k_{3}+k_{4}+k_{s})}=\frac{\langle J_1\varphi_2\varphi_{-\vec{s}}\rangle\langle\varphi_{\vec{s}}\varphi_3 J_4\rangle}{E}.
\fe
Here, $E=k_1+k_2+k_3+k_4$ is the total energy. Inspired by the method of obtaining $\Pi_{1,1}D_{uv}+\Pi_{1,0}\Delta_u$ (reviewed in section \ref{sec2}), we want to act some operators on the 3-pt correlator factors so that we can raise the spin of both the internal line and the external legs. A natural choice for the operator acting on the 3-pt correlator factors is $S^{++}_{ij}$, which raises the spin of two legs connected to the same point in the 3-pt case\footnote{One may wonder if we can use the operators like $(H_{12}+D_{11}D_{22}-2D_{12}D_{21})\ma{W}_{12}^{++}$. A simple calculation shows that this case is equivalent to the case we discuss now. The only difference is the value of some coefficients.}.

First we derive the result after acting $S^{++i}_{ij}$:
\ie
    S_{2,-s}^{++i}\langle J_1\varphi_2\varphi_{-s}\rangle&=-k^i_s\left(-\frac{2\vec{(k_1}\cdot\vec{\epsilon}_2) (\vec{k}_2\cdot\vec{\epsilon}_1)}{k_s(k_1+k_2+k_s)^2}+\frac{2(\vec{\epsilon}_1\cdot\vec{\epsilon}_2)}{k_1+k_2+k_s}\right)+\epsilon^i_2\frac{2(\vec{k}_2\cdot\vec{\epsilon}_1) }{k_1+k_2+k_s}\\
    S_{s3}^{++j}\langle\varphi_s\varphi_3 J_4\rangle&=-k^j_s \left(\frac{2 (\vec{k}_3\cdot\vec{\epsilon}_4) (\vec{k}_4\cdot\vec{\epsilon}_3)}{k_s (k_3+k_4+k_s)^2}-\frac{2(\vec{\epsilon}_3\cdot\vec{\epsilon}_4)}{k_3+k_4+k_s}\right)-\epsilon_3^j\frac{2 (\vec{k}_3\cdot\vec{\epsilon}_4)}{k_3+k_4+k_s}
\fe
Then, we can get 
\ie\label{cancel}
    &(\Pi_{1})_{ij}S_{2s}^{++i}\langle J_1\varphi_2\varphi_s\rangle S_{s3}^{++j}\langle\varphi_s\varphi_3 J_4\rangle\\
    &=-\frac{4(\vec{\epsilon}_2\cdot\vec{\epsilon}_3)(\vec{k}_2\cdot\vec{\epsilon}_1)(\vec{k}_3\cdot\vec{\epsilon}_4)}{(k_1+k_2+k_s)(k_3+k_4+k_s)}+\text{terms symmetric under $(1\leftrightarrow2)$ or $(3\leftrightarrow4)$}.
\fe
The first term in the second line is exactly $-ES^{++}_{23}k_{2}k_{3}D_{12}D_{43}F_{\Delta=2}$. There is a singularity at $\Delta=2$ in ``terms symmetric under $(1\leftrightarrow2)$ or $(3\leftrightarrow4)$". This will not bother us because if we sum over all the four terms of the second line of the equation \eqref{jjjj}, ``terms symmetric under $(1\leftrightarrow2)$ or $(3\leftrightarrow4)$" will be canceled. One may wonder if there will be terms with a factor $\Delta-2$ when we consider a general $\Delta$ and exactly cancel the singularity, since such terms may not satisfy the $(1\leftrightarrow2)$ and $(3\leftrightarrow4)$ symmetry. In other words, one may wonder if it is valid to take $\Delta=2$ first in our calculations. The answer is yes. Here we show some illuminating calculations.

From some CFT methods, one will find that the 3-pt function of a spin-1 current and two scalars can be determined up to a coefficient $f_{JOO}$:
\ie\label{ooj}
\la J_{\Delta_{1}}(x_{1})O_{\Delta_{2}}(x_{2})O_{\Delta_{3}}(x_{3})\ra=\left(\frac{\vec{x}_{21}\cdot \vec{\epsilon}_{1}}{x_{21}^{2}}-\frac{\vec{x}_{31}\cdot \vec{\epsilon}_{1}}{x_{31}^{2}}\right)\frac{f_{JOO}}{x_{23}^{\Delta_{2}+\Delta_{3}-\Delta_{1}+l}x_{31}^{\Delta_{3}+\Delta_{1}-\Delta_{2}-l}x_{12}^{\Delta_{1}+\Delta_{2}-\Delta_{3}-l}},
\fe
where $x_{ij}=|\vec{x}_{i}-\vec{x}_{j}|$ is the distance between the operator $O_{i}$ and $O_{j}$ at the boundary of dS spacetime. One can take the Fourier transformation to obtain the cosmological correlators in the momentum space. Thus we will find that we can get \eqref{ooj} in momentum space after acting some derivatives on the 3-pt correlators of 3 scalars (also in momentum space). More precisely,
\ie
\la J_{\Delta_{1}}(k_{1})O_{\Delta_{2}}(k_{2})O_{\Delta_{3}}(k_{3})\ra=&c_{1}(\vec{\epsilon}_{1}\cdot \vec{K}_{21})\la O_{\Delta_{1}}(k_{1})O_{\Delta_{2}+1}(k_{2})O_{\Delta_{3}}(k_{3})\ra\\
&-c_{2}(\vec{\epsilon}_{1}\cdot \vec{K}_{31})\la O_{\Delta_{1}}(k_{1})O_{\Delta_{2}}(k_{2})O_{\Delta_{3}+1}(k_{3})\ra,
\fe
where $c_{1}$ and $c_{2}$ denote the ratio of the coefficients of 3-pt correlators (e.g. $f_{JOO}/f_{OOO}$). After taking $\Delta_{1}=\Delta_{2}=2$, we have
\ie
\la J(k_{1})\varphi(k_{2})O_{\Delta}(k_{3})\ra=c_{1} (\vec{\epsilon}_{1}\cdot \vec{K}_{21})\la \varphi(k_{1})\phi(k_{2})O_{\Delta}(k_{3})\ra-c_{2}(\vec{\epsilon}_{1}\cdot \vec{K}_{31})\la \varphi(k_{1})\varphi(k_{2})O_{\Delta+1}(k_{3})\ra.
\fe
It is not hard to get the answer by Mathematica\footnote{here $c_{1}$ and $c_{2}$ are all included in the overall constant of the following equation, which we have omitted.}:
\ie\label{joo}
&\la J(k_{1})\varphi(k_{2})O_{\Delta}(k_{3})\ra\sim \\
&(\vec{\epsilon}_{1}\cdot \vec{k}_{2})\bigg[2^{\Delta -\frac{5}{2}} \Gamma (3-\Delta ) \Gamma \left(\Delta -\frac{3}{2}\right) \left(k_1+k_2\right){}^{\Delta -3} \, _2F_1\left(\frac{3-\Delta }{2},2-\frac{\Delta }{2};\frac{5}{2}-\Delta ;\frac{k_3^2}{\left(k_1+k_2\right){}^2}\right)\\
&+2^{\frac{1}{2}-\Delta } \Gamma \left(\frac{3}{2}-\Delta \right) \Gamma (\Delta ) k_3^{2 \Delta -3} \left(k_1+k_2\right){}^{-\Delta } \, _2F_1\left(\frac{\Delta }{2},\frac{\Delta +1}{2};\Delta -\frac{1}{2};\frac{k_3^2}{\left(k_1+k_2\right){}^2}\right)\bigg],
\fe
when we take $\Delta\rightarrow2$, the equation above will change to
\ie
\la J(k_{1})\varphi(k_{2})\varphi(k_{3})\ra\sim\frac{(\vec{\epsilon}_{1}\cdot \vec{k}_{2})}{k_{1}+k_{2}+k_{3}},
\fe
which matches with the result in \cite{Baumann:2020dch}. Then we can act $S^{++i}_{23}$ ($S^{++}_{23}=\sum_{i}\epsilon_{3}^{i}S^{++i}_{23}$) on $\la J(k_{1})\varphi(k_{2})O_{\Delta}(k_{3})\ra$, we have
\ie
S^{++i}_{23}&\la J(k_{1})\varphi(k_{2})O_{\Delta}(k_{3})\ra\\
&\sim(\Delta-1)\epsilon_{2}^{i}(\vec{\epsilon}_{1}\cdot \vec{k}_{2})2^{\Delta -\frac{5}{2}} \Gamma (3-\Delta ) \Gamma(\Delta -\frac{3}{2})k_{12}^{\Delta -3} \, _2F_1(\frac{3-\Delta }{2},2-\frac{\Delta }{2};\frac{5}{2}-\Delta ;k_r^2)\\
&+(\Delta-1)\epsilon_{2}^{i}(\vec{\epsilon}_{1}\cdot \vec{k}_{2})2^{\frac{1}{2}-\Delta } \Gamma (\frac{3}{2}-\Delta) \Gamma (\Delta ) k_3^{2 \Delta -3} k_{12}^{-\Delta } \, _2F_1(\frac{\Delta }{2},\frac{\Delta +1}{2};\Delta -\frac{1}{2};k_r^2)\\
&+k_{3}^{i}\bigg[(\vec{\epsilon}_1\cdot\vec{\epsilon}_2)2^{\Delta -\frac{5}{2}} \Gamma (3-\Delta ) \Gamma(\Delta -\frac{3}{2}) k_{12}^{\Delta -3} \, _2F_1(\frac{3-\Delta }{2},2-\frac{\Delta }{2};\frac{5}{2}-\Delta ;k_r^2)\\
&+(\vec{\epsilon}_1\cdot\vec{\epsilon}_2)2^{\frac{1}{2}-\Delta } \Gamma(\frac{3}{2}-\Delta) \Gamma (\Delta ) k_3^{2 \Delta -3}k_{12}^{-\Delta } \, _2F_1(\frac{\Delta }{2},\frac{\Delta +1}{2};\Delta -\frac{1}{2};k_r^2)\\
&+\frac{(\vec{\epsilon}_2\cdot \vec{k}_1) (\vec{\epsilon}_{1}\cdot \vec{k}_{2})}{\Delta -\frac{1}{2}}2^{-\Delta -\frac{1}{2}} \Delta  (\Delta +1) \Gamma(\frac{3}{2}-\Delta) \Gamma (\Delta ) k_3^{2 \Delta -3} k_{12}^{-\Delta -2} \, _2F_1(\frac{\Delta }{2}+1,\frac{\Delta +1}{2}+1;\Delta +\frac{1}{2};k_r^2)\\
&+\frac{(\vec{\epsilon}_2\cdot \vec{k}_1) (\vec{\epsilon}_{1}\cdot \vec{k}_{2})}{\frac{5}{2}-\Delta}2^{\Delta -\frac{5}{2}}(2-\frac{\Delta }{2}) (3-\Delta )\Gamma (3-\Delta ) \Gamma(\Delta -\frac{3}{2})k_{12}^{\Delta -5} \, _2F_1(\frac{5-\Delta }{2},3-\frac{\Delta }{2};\frac{7}{2}-\Delta ;k_r^2)\\
&+(\vec{\epsilon}_2\cdot \vec{k}_1) (\vec{\epsilon}_{1}\cdot \vec{k}_{2})2^{\frac{1}{2}-\Delta } (2 \Delta -3) \Gamma(\frac{3}{2}-\Delta) \Gamma (\Delta ) k_3^{2 \Delta-5}k_{12}^{-\Delta } \, _2F_1(\frac{\Delta }{2},\frac{\Delta +1}{2};\Delta -\frac{1}{2};k_r^2))\bigg],
\fe
To simplify our notation, we introduce two new variables: $k_{12}=k_1+k_2$ and $k_r= k_3/(k_1+k_2)$. Here particle 3 will be treated as the propagating particle. It is not hard to find that things are the same as before: the singularity at $\Delta=2$ will be canceled! In fact, there will be no term proportional to $\Delta-2$ in $S^{++i}_{23}\la J\varphi O_{\Delta}\ra$, which means taking the limit $\Delta\rightarrow 2$ before acting $S^{++i}_{ij}$ is valid in this case.

So far, we have demonstrated how to construct $\la JJJJ\ra$ and the consistency with the weight-shifting perspective. In fact, there may be deeper reasons for such a construction. Inspired by the traditional uplifting method \cite{Baumann:2021fxj,Bonifacio:2022vwa,Li:2022tby,Mei:2023jkb}, we find that the equation \eqref{jjjj} can be obtained by some replacement of the flat 4-pt gluon amplitude and will show this fact in the following discussion.

Now we write down the $s$-channel terms of the flat 4-pt gluon amplitude appearing in \cite{Cheung:2017ems}:
\ie
&-\frac{1}{2S}\{(\vec{\epsilon}_1\cdot\vec{\epsilon}_2)(\vec{\epsilon}_3\cdot\vec{\epsilon}_4)(-U+T)-4(\vec{\epsilon}_{2}\cdot\vec{\epsilon}_{3})(\vec{k}_{3}\cdot\vec{\epsilon}_{4})(\vec{k}_{2}\cdot\vec{\epsilon}_{1})+4(\vec{\epsilon}_{2}\cdot\vec{\epsilon}_{4})(\vec{k}_{4}\cdot\vec{\epsilon}_{3})(\vec{k}_{2}\cdot\vec{\epsilon}_{1})\\
&+4(\vec{\epsilon}_{1}\cdot\vec{\epsilon}_{3})(\vec{k}_{3}\cdot\vec{\epsilon}_{4})(\vec{k}_{1}\cdot\vec{\epsilon}_{2})-4(\vec{\epsilon}_{1}\cdot\vec{\epsilon}_{4})(\vec{k}_{4}\cdot\vec{\epsilon}_{3})(\vec{k}_{1}\cdot\vec{\epsilon}_{2})\\
&+4(\vec{\epsilon}_1\cdot\vec{\epsilon}_2)[(\vec{k}_{1}\cdot\vec{\epsilon}_{3})(\vec{k}_{2}\cdot\vec{\epsilon}_{4})-(\vec{k}_{2}\cdot\vec{\epsilon}_{3})(\vec{k}_{1}\cdot\vec{\epsilon}_{4})]+4(\vec{\epsilon}_3\cdot\vec{\epsilon}_4)[(\vec{k}_{3}\cdot\vec{\epsilon}_{1})(\vec{k}_{4}\cdot\vec{\epsilon}_{2})-(\vec{k}_{4}\cdot\vec{\epsilon}_{1})(\vec{k}_{3}\cdot\vec{\epsilon}_{2})]\}
\fe
where $T=-(\vec{k}_{1}+\vec{k}_{4})^{2}$ and so as $U$, $S$, which are the Mandelstem varieties in flat spacetime. Note that in flat spacetime we have $k^{2}=0$ for massless particles. After the following replacement:
\ie
\vec{\epsilon}_{i}\cdot\vec{\epsilon}_{j}&\rightarrow S^{++}_{ij}\\
\vec{\epsilon}_{i}\cdot \vec{k}_{j}&\rightarrow k_{j}D_{ij} \ (\text{$i,j$ on the same side of the channel})\\
T-U=-S-2U&\rightarrow \Pi_{1,1}D_{uv}+\Pi_{1,0}\Delta_{u}\\
2(\vec{\epsilon}_{i}\cdot\vec{k}_{j})(\vec{\epsilon}_{k}\cdot\vec{k}_{l})&\rightarrow \vec{\epsilon}_{i}\circ\vec{\epsilon}_{k}D_{uv} \ (\text{$i,j$ (also $k,l$) on the different sides of the channel})\\
-\frac{1}{2S}&\rightarrow F_{\Delta=2},
\fe
we will reproduce \eqref{jjjj}. In this replacement, $k_s^2$ corresponds to $D_{uv}$, and $-S-2U$ corresponds to the operator that raises the spin of the internal line. This can be explained by comparing the ``polarization sum" $\ma{P}_{1}=\Pi_{1,1}s^2-\Pi_{1,0}(k_1+k_2+k_s)(k_3+k_4+k_s)$ in the factorization method (the definition of $\ma{P}_{1}$ and more about cosmological bootstrap can be found in \cite{Baumann:2020dch}) with the spin-raising operators $\Pi_{1,1}D_{uv}+\Pi_{1,0}\Delta_{u}$. Note that $-S-2U$ is the flat limit of $\ma{P}_{1}$.

From \eqref{jjjj}, we find that we cannot simply get $\la JJJJ\ra$ from $\la JOJO\ra$ or $\la JJOO\ra$, but we need to consider all of those and finally express the correlators by acting some operators on the seed $F_{\Delta=2}$. It means the trial for seeking the inverse of the unifying relation fails, like the flat case. 

\section{Outlooks}
In this paper, we have reviewed two ways, weight-shifting operators and unifying relations, to get other cosmological correlators from the given one and made some new comments. Then we showed that the trial for seeking the inverse of unifying relations fails. However, we have found a ``weight-shifting uplifting" method similar to the traditional uplifting method for dS correlators in our special case, which is a hint for applying weight-shifting operators to higher point correlators. There are several open problems inspired by this work:
\begin{enumerate}
    \item Can we find other weight-shifting operators to raise the spin of the internal lines? In section \ref{sec4}, we have constructed a new way to raise the spin of the internal line for the special case we met. Maybe there are some other similar cases, and they will lead to some new allowed structures. 
    \item Can this prescription be generalized to BG currents so that we can get $n$-pt results by BG recursions? For weight-shifting operators, it is very hard to use them to deal with higher-point correlators. However, from the structure of BG currents in dS spacetime, even $n$-pt cosmological correlators have some features similar to the flat case. This may imply that the uplifting way may lead to a higher-point correspondence between the cosmological correlators and the flat amplitudes.
    \item Can we do the same thing for the graviton correlators\cite{Bonifacio:2022vwa,Albayrak:2023jzl}? Maybe we need to prove unifying relations for the gravity theory. However, in flat spacetime, the gravity theory we deal with is the extended gravity theory, which includes dilatons and B-fields. Recently, there are some progress in the double copy of the cosmological correlators\cite{Armstrong:2023phb,Mei:2023jkb}, which may help us to find out the ``weight-shifting uplifting" for graviton correlators in dS spacetime.
\end{enumerate}

There are still many things unknown in cosmological correlators. The interesting problems above are just a few of them. Solving them will deepen our understanding of cosmological correlators and, moreover, our universe. We hope we can solve some of them in future work.

\section*{Acknowledgements}
We would like to thank Hayden Lee for enlightening discussions, and Jiajie Mei for discussions and comments on our draft. YT is partly supported by National Key R\&D Program of China (NO. 2020YFA0713000). 
\appendix
\section{More on weight-shifting operators}\label{appA}
In this appendix, we will give more comments on weight-shifting operators. In the context of cosmological applications, expressing correlators in momentum space is highly convenient. However, we need to keep in mind that in the Poincaré patch of de Sitter space (dS$_4$), we lack a timelike Killing vector, which prevents us from performing a Fourier transformation with respect to the time direction. Instead, the cosmological correlator in momentum space can be expressed as an integral over conformal time, often referred to as the seed integral. Subsequently, the weight-shifting operators can be represented as differential operators with respect to momentum, acting on the correlators in momentum space.

One effective approach to identifying the appropriate weight-shifting operator is by examining the seed integral. In essence, the cosmological correlator can be expressed as an integral over conformal time, incorporating both bulk-to-bulk and bulk-to-boundary propagators. Although the integrand of various correlators may differ, we can establish a connection between these integrands using differential operators in terms of momentum. Since the derivative with respect to momentum can be commuted with the time integral, we can establish correlations among correlators of diverse theories using these differential operators, which precisely correspond to the weight-shifting operators.

To establish a clearer connection between different theories, we can begin by explicitly formulating the integral for various correlator expressions. 
The starting point of this work is the well-established 3-point scalar correlator with a general conformal weight in momentum space. Let us begin by presenting the seed integral for this particular correlator\cite{Bzowski:2013sza}:
\begin{equation}\label{eq:seedintegral}
    \langle\ma{O}_1\ma{O}_2\ma{O}_3\rangle=k_1^{\Delta_1-\frac{3}{2}}k_2^{\Delta_2-\frac{3}{2}}k_3^{\Delta_3-\frac{3}{2}}\int_0^{\infty}dz z^{\frac{1}{2}}K_{\Delta_1-\frac{3}{2}}(k_1 z)K_{\Delta_2-\frac{3}{2}}(k_2 z)K_{\Delta_3-\frac{3}{2}}(k_3 z).
\end{equation}
It is worth noting that the expression provided above neglects an overall normalization coefficient. Additionally, the symbol $K$ represents the Bessel-$K$ function. Interestingly, for certain special conformal dimensions $\Delta$, the integral can be expressed using elementary functions. In the context of cosmological applications, our primary focus is on conformally coupled and massless scalars, which have conformal dimensions of $\Delta=2$ and $\Delta=3$, respectively. Remarkably, for these two conformal dimensions, the integral can be simplified precisely to elementary functions. For example, the 3-point correlator for the conformally coupled scalar can be expressed as follows:
\begin{equation}
    \langle\varphi\varphi\varphi\rangle=\log\left(\frac{K}{\mu}\right)
\end{equation}
where $K=k_1+k_2+k_3$ is the total momentum and $\mu$ represents an energy scale that can be chosen arbitrarily. It is important to mention that the logarithm arises due to the IR divergence in the integral. This divergence becomes evident from the bulk perspective, where the seed integrand for $\varphi\varphi\varphi$ is proportional to $e^{-Kz}/z$, exhibiting divergence in the infrared limit. The apparent dependence on the energy scale violates dilatation symmetry, which is why it is commonly referred to as the anomalous term from the boundary perspective.

Now, we aim to introduce the weight-shifting operators and demonstrate their effects using the seed integral. To provide a clearer illustration, let us introduce the simplest weight-shifting operators $\mathcal{W}^{--}_{12}$ as an example, which decrease the conformal dimension by one unit at each point (in this case, particle 1 and particle 2).
Back to the position space, the seed integral can be expressed as:
\begin{equation}
    \langle\ma{O}_1(\vec{x}_1)\ma{O}_2(\vec{x}_2)\ma{O}_3(\vec{x}_3)\rangle=\frac{f_{123}}{\left|\vec{x}_1-\vec{x}_2\right|^{\Delta_1+\Delta_2-\Delta_3}\left|\vec{x}_2-\vec{x}_3\right|^{\Delta_2+\Delta_3-\Delta_1}\left|\vec{x}_3-\vec{x}_1\right|^{\Delta_3+\Delta_1-\Delta_2}},
\end{equation}
where $f_{123}$ is an overall constant that is not crucial for our discussion and thus we can disregard such coefficients in the subsequent analysis. In order to obtain the 3-point scalar with conformal dimensions $\Delta_1$, $\Delta_2$, and $\Delta_3$ in momentum space, we can perform a Fourier transformation:
\begin{equation}
    \langle\ma{O}_1\ma{O}_2\ma{O}_3\rangle=\int d\vec{x}_1d\vec{x}_2d\vec{x}_3\frac{e^{i\vec{p}_1\cdot\vec{x}_1}e^{i\vec{p}_2\cdot\vec{x}_2}e^{i\vec{p}_3\cdot\vec{x}_3}}{\left|\vec{x}_1-\vec{x}_2\right|^{\Delta_1+\Delta_2-\Delta_3}\left|\vec{x}_2-\vec{x}_3\right|^{\Delta_2+\Delta_3-\Delta_1}\left|\vec{x}_3-\vec{x}_1\right|^{\Delta_3+\Delta_1-\Delta_2}}
\end{equation}
The derivative with respect to momentum can be interchanged with the integral over position variables $\vec{x}_1$, $\vec{x}_2$, and $\vec{x}_3$. Therefore, we can apply the weight-shifting operator $\mathcal{W}^{--}_{12}$ to the correlator, placing it in front of the integral. Subsequently, we insert it into the integrand, and these differential operators act solely on the exponential term involving momentum. As a result, we obtain the 3-point scalar correlator with the conformal dimensions $\Delta_1$ and $\Delta_2$, both reduced by one unit:
\begin{equation}
\begin{split}
    \langle\ma{O}_{\Delta_1-1}(\vec{k}_1)\ma{O}_{\Delta_2-1}(\vec{k}_2)\ma{O}_{\Delta_3}(\vec{k}_3)\rangle&=\int d\vec{x}_1d\vec{x}_2d\vec{x}_3\frac{\left|\vec{x}_1-\vec{x}_2\right|^{2}e^{i\vec{k}_1\cdot\vec{x}_1}e^{i\vec{k}_2\cdot\vec{x}_2}e^{i\vec{k}_3\cdot\vec{x}_3}}{\left|\vec{x}_1-\vec{x}_2\right|^{\Delta_1+\Delta_2-\Delta_3}\left|\vec{x}_2-\vec{x}_3\right|^{\Delta_2+\Delta_3-\Delta_1}\left|\vec{x}_3-\vec{x}_1\right|^{\Delta_3+\Delta_1-\Delta_2}}\\
    &\sim\int d\vec{x}_1d\vec{x}_2d\vec{x}_3\frac{\ma{W}^{--}_{12}e^{i\vec{k}_1\cdot\vec{x}_1}e^{i\vec{k}_2\cdot\vec{x}_2}e^{i\vec{k}_3\cdot\vec{x}_3}}{\left|\vec{x}_1-\vec{x}_2\right|^{\Delta_1+\Delta_2-\Delta_3}\left|\vec{x}_2-\vec{x}_3\right|^{\Delta_2+\Delta_3-\Delta_1}\left|\vec{x}_3-\vec{x}_1\right|^{\Delta_3+\Delta_1-\Delta_2}}\\
    &=\ma{W}^{--}_{12}\langle\ma{O}_{\Delta_1}(\vec{k}_1)\ma{O}_{\Delta_2}(\vec{k}_2)\ma{O}_{\Delta}(\vec{k}_3)\rangle
\end{split}
\end{equation}
The notation $\sim$ means that we omit some overall constants. Such constants can be restored by the explicit calculation on the cosmological correlator side. The aforementioned calculations indicate that we can construct the weight-shifting operators by comparing the correlator of our interest with the 3-point correlator in position space and subsequently transforming it to momentum space.  Remarkably, this construction ensures that the weight-shifting operators possess the same quantum numbers for both spin and conformal dimensions. Henceforth, we will directly present the weight-shifting operator without explicitly demonstrating the construction procedure. However, it is essential to bear in mind that the construction procedure is nearly identical to what we previously employed in constructing $\mathcal{W}^{--}_{12}$. To learn more about the construction procedure, see \cite{Baumann:2019oyu}, where they use the embedding space formalism.

\section{4-pt Scalar Seed correlators}
In this appendix, we introduce the 4-point scalar seed correlators. It is important to note that the 4-point scalar seed integral differs from the 3-point correlators, as it involves two conformal time integrals, adding complexity to the time ordering. Our main focus in the following discussions will be on the scalar exchange solution

The conventional approach to compute vacuum expectation values in a time-dependent background is through the Schwinger-Keldysh (SK) formalism\cite{Weinberg:2005vy}. For instance, considering the interaction $g\varphi^2\sigma$, where $\sigma$ is a scalar field with an arbitrary conformal dimension, its seed integral can be expressed as follows:
\begin{equation}\label{eq:4-ptseed}
    F=-\frac{g^2}{2}\sum_{\mb{ab}}\int_{-\infty}^0\frac{d\eta}{\eta^2}\frac{d\eta'}{\eta'{}^2}e^{i\mb{a}k_{12}\eta}e^{i\mb{b}k_{34}\eta'}G_{\mb{ab}}(k_s;\eta,\eta'),
\end{equation}
where $G_{\mb{ab}}(k_s;\eta,\eta')$ represents the bulk-to-bulk propagator for the massive scalar $\sigma$, where $\mb{a,b}=\pm$ denote the SK indices. And $\eta$ represents the conformal time, and $g$ denotes the coupling constant between the fields $\varphi$ and $\sigma$. The specific form of the propagator $G_{\mb{ab}}(k_s;\eta,\eta')$ depends on the particular model under consideration, and we will not delve further into its details in this study \cite{Arkani-Hamed:2018kmz}.


For our specific application, we can provide the explicit expression for the seed integral when the exchanged particle $\sigma$ has a conformal dimension $\Delta=2$. This solution can be derived using both the bulk and boundary perspectives. Here, we present the result without going into the detailed derivation. For a more comprehensive understanding, interested readers are encouraged to refer to previous papers such as \cite{Arkani-Hamed:2015bza, Arkani-Hamed:2018kmz}:
\begin{equation}
   F_{\Delta_{\sigma}=2}=\frac{1}{2k_s}\left[\te{Li}\left(\frac{k_{34}-k_s}{E}\right)+\te{Li}\left(\frac{k_{12}-k_s}{E}\right)+\log\left(\frac{k_{12}+k_s}{E}\right)\log\left(\frac{k_{34}+k_s}{E}\right)\right].
\end{equation}
In the above expression, we have introduced the following notations: $E=k_1+k_2+k_3+k_4$ as the total energy, and $k_{12}=k_1+k_2$ and $k_{34}=k_3+k_4$ for convenience. Additionally, the symbol ``$\te{Li}$" denotes the di-logarithm function. It is worth mentioning that even though the boost symmetry is broken, we can also derive the seed integrals and the corresponding weight-shifting operators \cite{Pimentel:2022fsc,Wang:2022eop}.

\bibliographystyle{JHEP}
\bibliography{WS}

\end{document}